\documentstyle[11pt]{article}
\newcommand{\blankline}{\vskip .3cm}
\newcommand{\f}{\begin{equation}}
\newcommand{\ff}{\end{equation}}
\newcommand{\bea}{\begin{eqnarray}}
\newcommand{\eea}{\end{eqnarray}}

\begin{document}
\centerline{\LARGE  Strings as perturbations}
\centerline{\LARGE  of evolving spin-networks}
\blankline
\rm
\centerline{Lee Smolin}
\blankline
\blankline
\centerline{Center for Gravitational Physics and 
Geometry}
\centerline{\it Department of Physics}
\centerline {\it The Pennsylvania State University}
\centerline{\it University Park, PA, USA 16802}
 \vfill
\centerline{December 28, 1997}
\vfill
\centerline{ABSTRACT}
A connection between non-perturbative formulations of quantum
gravity  and perturbative string theory
is exhibited, based on a formulation of the non-perturbative dynamics 
due to Markopoulou.
In this formulation the dynamics of spin network states and their
generalizations is described in terms
of histories which have discrete analogues of the 
causal structure and many fingered
time of Lorentzian spacetimes.  Perturbations of these histories 
turn out to be described in  terms of spin systems defined on 2-dimensional 
timelike surfaces embedded in the discrete spacetime. When the 
history has a classical limit which is Minkowski spacetime,  
the action of the perturbation theory is given to leading order
by the spacetime area 
of the surface, as in bosonic string theory.  This map between a 
non-perturbative
formulation of quantum gravity and a 1+1 dimensional theory 
generalizes to
a large class of theories in which the group $SU(2)$ is extended 
to any quantum
group or supergroup. It is argued that a necessary 
condition for the non-perturbative theory to have a good
classical limit is that the resulting 1+1
 dimensional theory
defines a consistent and stable perturbative
string theory.  
\blankline
email address:  smolin@phys.psu.edu 
\eject
\section{Introduction}

There are two approaches to quantum gravity that have made 
significant progress in the last ten years.  In the perturbative
or, more precisely, the background dependent, regime, string theory
provides the only consistent description of the interaction of
gravitons with other degrees of freedom.  Moreover, there are
good reasons  to believe that whatever the true quantum
theory of gravity is, string theory will describe its perturbative
limit.  Among these is that it  resolves a paradox,
which is how to have a theory which has a physical cutoff at a
fixed length scale while remaining Lorentz 
invariant\cite{lenny-holo}.

At the same time, while there are many results that point to the
existence of a non-perturbative, background independent theory
that unifies the various perturbative string theories, we do not yet 
know the form of that theory (despite some very interesting
proposals\cite{matrix,dvv,horava}). Given the fact that 
S duality\cite{Sduality} and mirror 
symmetry\cite{mirror} strongly suggest that the non-perturbative theory 
cannot be expressed in terms of the embeddings
of strings, membranes or anything else in a fixed manifold, from
what mathematical elements
could non-perturbative string theory be constructed?
If we remove manifolds from  
the quantum theory, one is left only with representation theory and
combinatorics.  

Could $\cal M$ theory be constructed
from only representation theory and combinatorics?  While the
answer is not known, we may note that 
quantum general relativity has been 
constructed
\cite{lp1,volume1,ham1,rigorous,qsd,carlo-review,future}
and it is almost of this form.  The states of the theory are labeled by
spin networks\cite{roger-sn,volume1}, which 
are constructed from combinatorics and
the representation theory of $SU(2)$.  The local 
operators of the theory may be expressed in terms of finite
combinatorial operations on these 
states\cite{spain,volume1,ham1}.  The result is a 
combinatorial
picture of quantum geometry in which 
areas\cite{spain,volume1}, volumes \cite{spain,volume1,othervolume} and
lengths\cite{thomas-lengths} are discrete and have computable 
spectra\footnote{Quantum general relativity is not completely
combinatorial because 
the states of those theories are labeled by the
embeddings of the spin networks in a prior three manifold
$\Sigma$, up to diffeomorphisms.  However, the local operators
that have so far been studied do not depend on the embedding
information.  To arrive at a purely 
combinatorial picture of quantum geometry the embedding should
be reconstructed from combinatorial information.   This may
be possible, given an appropriate non-embedded
extension of the spin network
states, as discussed in \cite{fotini1,fotinilee2}.}.

At the Hamiltonian level, quantum general relativity exists
as a sensible quantum theory\cite{qsd}; 
the only problem with it is that,
as is the case for example for random surface theories away from
their critical points, it seems to lack a continuum limit which
describes massless particles moving in a Lorentzian 
spacetime\cite{trouble1}.
The evidence is that what has been constructed, at least
so far, is the ultralocal, or $c \rightarrow 0$ limit 
of the theory\cite{trouble1,trouble2}.

It is then natural to ask whether there might be some extension of
the spin network states of quantum general relativity that may
describe a non-perturbative string theory.  The kinds of extensions
that might be explored include the following:  1) extend the group
whose representation theory labels the spin networks from $SU(2)$,
which is related to symmetries of $3+1$ dimensional spacetime
to other groups relevant for string theory, 2) add supersymmetry,
3)  add labeled surfaces corresponding to p-form gauge 
theories,  4)  make the theory completely background independent
and combinatorial
by removing the embedding of the spin networks 
in fixed manifolds  
5) extend the dynamics
from that given by general relativity to a larger set, to be
constrained only by some appropriate notion of local causality.

The study of all of these extensions is in progress.  Supersymmetry
seems to be naturally incorporated, 
as in \cite{superloop}, as are p-form
fields\cite{diffeomorphic}.  New general forms of the 
dynamics have been
explored, in both the Euclidean 
\cite{mike,carlomike,louis,louisjohn,baez,barberi} and  
Lorentzian \cite{fotinilee1,fotini1,fotinilee2}
cases.   How to extend the group from $SU(2)$
to any quantum group or Hopf algebra $G_q$, while at the same
time dropping the dependence on embedding, has been understood
in \cite{fotinilee2}\footnote{Quantum deformation to
$SU(2)_q$ is required to incorporate a cosmological constant
with $q=e^{2\pi i/k+2}$ with 
$k = 6\pi/G^2 \Lambda$ \cite{linking,qdeform}. }.  

The next question is how should one test whether such an extension
 might yield a non-perturbative formulation of $\cal M$
theory.  (Or to put it in the 
native dialect, if a completely compactified version of $\cal M$ 
theory walked in the door, how would we recognize it?)  One
necessary condition is that the theory must have a classical
limit in which some Lorentzian manifolds turn out to be a good
approximate description and general relativity is approximately
true.  A second necessary condition is that perturbation theory 
around this limit must be described in terms of the interactions of
strings and branes in that manifold.  

The problem of the classical limit is hard as it is a problem
in critical phenomena.  As the causal structure itself is dynamical
this problem seems to be more analogous to non-equilibrium critical
phenomena such as directed percolation than it is to second
order phase transitions\cite{fotinilee1}.   Here we will assume that
this problem has been solved and tackle the second problem, that
of the perturbation theory around a classical limit.  We will 
study a class of theories which are characterized
by the choice of a quantum group or superalgebra
$G_q$ and a set of evolution amplitudes $\cal A$ (to be
characterized below.)  We will show that these theories
have the following properties:  

\begin{enumerate}

\item{}Their perturbation theory
is described  
in terms of a $1+1$ dimensional spin system defined
on a timelike surface embedded in the Lorentzian spacetime
$(M,g)$ that arises in the classical limit of the theory.

\item{}If that spacetime $(M,g)$ is Minkowski spacetime
and the group $G=SU(2)_q$, the
effective action for the perturbation associated with a surface
$X^\mu (\sigma, \tau )$ is to leading order 
proportional to the Nambu action
of bosonic string theory, which is its area.

\item{}A necessary condition that one of this
class of theories has a good classical
limit is that the induced $1+1$ dimensional
theory describes a consistent string theory in $3+1$
dimensions.

\end{enumerate}

To find these results we work with a form of the dynamics
of spin networks proposed by Markopoulou, which has built
in local causality\cite{fotini1}.  While we have recently extended
that formalism to a large class of theories in which the spin
networks are extended to states defined in terms of the
conformal blocks of a rational conformal field 
theory\cite{fotinilee2}, we
will work here with a restricted set of the states of these
theories which can be described in terms of labeled triangulations
of some manifold.  The reason is that in the absence of a real
understanding of the classical limit this allows us to sidestep
some questions associated with that limit.  We will also work
here with the group $SU(2)_q$, although the extension to
any group or Hopf algebra $G_q$ is straightforward.

Before starting, we note that we are not claiming to 
have shown that the theories we discuss here are non-perturbative
string theories. Nor have we given sufficient conditions for
them to be.  However, we may note that a dynamical theory of
the $(p,q)$ strings, which are a large set of $BPS$ states
may be naturally formulated in this 
framework\cite{pqstring}.  

In the next two sections we summarize the kinematical and
dynamical framework for quantum theories of gravity with
which we will work \cite{fotini1}.  
In section 4 we describe the perturbation theory and show
how a $1+1$ theory is constructed to describe the perturbations.

\section{Summary of quantum spatial  geometry in the dual picture}

In the dual picture described by Markopoulou\cite{fotini1} the 
space of states of a non-perturbative quantum theory of gravity
based on a group $G_q$ is constructed as follows.  

We begin with a  three manifold $\Sigma$,
to which we will associate a space of states ${\cal H}^\Sigma$ 
Consider a simplicial decomposition $\cal T$
of $\Sigma$.  $\cal T$ is labelled as follows.
Each 2-dimensional 
face, $f$,  in the triangulation $\cal T$ is 
labelled by a representation $j^{(f)}$ of $G_q$.   Each tetrahedron 
$\tau\in {\cal T}$ 
is labelled by an intertwiner map $\mu^{(\tau )} \in 
V_{ijkl}^{(\tau)}$ 
for the four representations on the faces of the tetrahedron. 

We may then denote the labelled triangulations as states
 $|{\cal T},\{ j \} , \{ \mu \} \rangle$ which we will require
to form an orthonormal basis of
${\cal H}^\Sigma$
The inner product on ${\cal H}^\Sigma$ is defined as
\f
\langle
{\cal T},\{ j \} , \{ \mu \} |{\cal T}^\prime ,\{ j' \}  , \{ \nu \} 
\rangle
= \delta_{{\cal T}{\cal T}^\prime}\delta_{\{ j \}\{ j' \}}
\prod_{\tau \in {\cal T}} 
\langle\mu^{(\tau )} |\nu^{(\tau )}\rangle_{V^{(\tau )}}
\label{ip}
\ff
where $\mu^{(\tau)}, \nu^{(\tau)}\in {\cal V}_{j_1 j_2 j_3 j_4}^{(\tau)}$, and 
$\langle\ |\ \rangle_{{\cal V}^{(\tau)}}$ is the natural inner product 
in ${\cal V}_{j_1 j_2 j_3 j_4}^{(\tau)}$. 
Thus, the states are orthogonal unless there is an isomorphism of
one triangulation to the other that preserves the labels on the faces.
If $I$ labels the tetrahedra $\tau\in{\cal T}$, then
given a basis $\{\mu^\alpha\}_I$ in the space
of intertwiners ${\cal V}_{j_1 j_2 j_3 j_4}^{I}$ the states 
$|{\cal T}^\prime ,\{ j\}  , \{ \mu^\alpha \}_I 
\rangle$ give an orthonormal basis of ${\cal H}\Sigma$.

In the $SU(2)$ case,  area\cite{spain,volume1} and 
volume\cite{spain,volume1,othervolume} operators 
may then be used to assign
areas to the faces and volume operators to each space
of intertwiners in each tetrahedron
\footnote{We 
may note that these state spaces ${\cal H}^\Sigma$ 
are subspaces of  the full set of non-embedded states
${\cal H}_{G_q}$ defined in \cite{fotinilee2}.  The correspondence
is constructed as follows.  
To each triangulation $\cal T$ we may associate a
dual 2-surface ${\cal S}[{\cal T}]$ by the  
following procedure.  Each tetrahedron $\tau_I\in{\cal T}$
is mapped to a 4-punctured sphere $B_I$, so that
the surface of $B_I$ has the same orientation as 
the surface of $\tau_I$. 
Two spheres
$B_I$ and $B_J$ are joined at a puncture when that puncture corresponds
to a face shared between the tetrahedra $\tau_I$ and $\tau_J$.
The punctured is labelled by the representation on that face. 
Given a state $|{\cal T} , \{ j\} , \{ \mu \}\rangle$ we then
get a state  $|{\cal S} [{\cal T}], \{ j \} , \{ \mu \}\rangle$ by
transferring the representations and interwiners according to
the construction.  This establishes a map
from ${\cal H}^\Sigma$ into a subspace
of ${\cal H}_{G_q}$. In the case $G=SU(2)$ these labeled two surfaces
are equivalent to labeled graphs, i.e. to spin networks.}.
 
\section{Summary of causal evolution in the dual picture}

According to the proposal of \cite{fotini1}
the states of $\cal S$ evolve according to a distinct set
of rules.  Any state $\Gamma_0$ may evolve to one of 
a finite number of possible successor states $\Gamma_0^I$.
Each $\Gamma^I_0$ is derived from $\Gamma_0$ by
application of one of four possible moves, called
Pachner moves\cite{pachner}.  These moves modify the state $\Gamma_0$
in a local region involving one to four adjacent
tetrahedra.

Consider
any subset of $\Gamma$ consisting of $n$ adjacent tetrahedra, 
where $n$ is between $1$ and $4$, which make up $n$ sides of a 
four-simplex $S_4$.  Then there is an evolution rule by which those $n$
tetrahedra are removed, and replaced by the other $5-n$ tetrahedra
in the $S_4$.  This is called a Pachner move.  The different possible
moves are called $n \rightarrow (5-n)$ moves 
(Thus, there are $1\rightarrow 4$, 
$2 \rightarrow  3$ etc. moves.) 
The new tetrahedra must be labeled, by 
new representations $j$ and intertwiners $k$.  
For each move there are $15$ labels involved, 
$10$ representations on the faces 
and $5$ intertwiners on the tetrahedra.  This is because
the labels involved in the move are exactly those of the
four simplex $S_4$.  
For each $n$  there is then an amplitude
${\cal A}_{n \rightarrow 5-n}$ that is a function of the $15$ labels.  A
choice of these amplitudes for all possible labels, for the four 
cases $1 \rightarrow 4 ,....,4\rightarrow 1$, then 
constitutes a choice of the dynamics
of the theory.

The application of one of the possible Pachner moves
to $\Gamma_0$, together with a choice of the possible
labelings on the new faces and tetrahedra the move
creates, results in a new spin network state $\Gamma_1$.
This differs from $\Gamma_0$ just in a region which
consisted of between $1$ and $4$ adjacent tetrahedra.
The process may be continued a finite number of times
$N$, to yield successor  states 
$\Gamma_2, ... \Gamma_N$.

Any particular set of $N$ moves beginning with a state
$\Gamma_0$ and ending with a state
$\Gamma_N$ defines a four dimensional 
combinatorial structure, which we will call a
{\it history}, $\cal M$ from $\Gamma_0$ to $\Gamma_N$.
Each history consists of $N$ combinatorial four simplices.
We will require that every tetrahedra in the real triangulation
has been subject to at least one move.  This means that
the boundary of $\cal M$,  which is a set of tetrahedra,  
falls into two connected sets so that
$\partial {\cal M} = \Gamma_0 \cup \Gamma_N$.  
All tetrahedra not in the boundary of $\cal M$ are
contained in exactly two four simplices of $\cal M$.

Each history $\cal M$ is a causal set, whose structure
is determined as follows.  The tetrahedra of each
four simplex, $S_4$ of $\cal M$ are divided into two
sets, which are called the past and the future set.
This is possible because each four simplex contains
tetrahedra in two states $\Gamma_i$ and $\Gamma_{i+1}$
for some $i$ between $0$ and $N$.
Those in $\Gamma_i$ were in the group that were wiped
out by the Pachner move, which were replaced by those
in $\Gamma_{i+1}$.  Those that were wiped out are
called the past set of that four simplex, the new ones,
those in $\Gamma_{i+1}$ are called the future set.
With the exception of those in the boundary, every
tetrahedron is in the future set of one four simplex
and the past set of another.

The causal structure of $\cal M$ is then defined as follows.
The tetrahedra of $\cal M$ make up a causal set
defined as follows\footnote{A causal set is a set with
a partial order, with no closed timelike loops\cite{rafael}.}.
Given two tetrahedra $T_1$ and $T_2$ in $\cal M$, we
say $T_2$ is to the future of $T_1$ (written $T_2 > T_1$)
if there is a sequence of causal steps that begin on
$T_1$ and end on $T_2$.  A causal step is a step from
a tetrahedron which is an element of the past set of
some four simplex, $S_4$ to any tetrahedron which is
an element of the future set of the same four simplex.
By construction, there are no closed causal loops, so the
partial ordering gives a causal set.

This theory then falls partly within the causal set
formulation of discrete quantum spacetime proposed
by Sorkin and collaborators \cite{rafael} and 
't Hooft\cite{thooft-causal}.
However, they have additional
structure.
Among these is the fact that
each history $\cal M$ may also be foliated by a number of
spacelike slices $\Gamma$.  A spacelike slice of $\cal M$
is a connected set of tetrahedra of $\cal M$ that 1)
constitute some $\Gamma \in {\cal S}$ (so it is dual
to a 4 valent spin network with no free ends) and
2) no two of whose elements are causally related.

Each $\Gamma_i$ in the original construction of
$\cal M$ constitutes a spacelike slice of $\cal M$.
But there are also many other spacelike slices
in $\cal M$ that are not one of the $\Gamma_i$.
In fact, given any spacelike slice $\Gamma$ in
$\cal M$ there are a large, but finite, number of
slices which are differ from it by the application
of one Pachner move.  Because of this, there
is in this formulation a discrete analogue of the
many fingered time of the canonical picture
of general relativity.  

The dynamics is then to be specified by the choice of the
amplitudes ${\cal A}[{\cal M}]$.  By locality these should
be products of amplitudes for each evolution step,
\f
{\cal A}[{\cal M}]=\prod_{I=1}^N {\cal A}_I[j,k,c]
\ff
where we have indicated that an amplitude  
${\cal A}_I[j,k,c]$ is associated with each causal step, which
is realized by the action of a four simplex.  It may in general
depend on the spins on the faces, $j$, the intertwiners on the
tetrahedra,
$k$ and the causal structure (division of the
tetrahedra into a future and past set, $c$.)

The transition amplitude 
from an initial state $\Gamma_0$ to a final state
$\Gamma_f$ is given by a  sum over all histories $\cal M$
that evolve $\Gamma_0$ to  
$\Gamma_f$.  
\f
{\cal A}[\Gamma_0 \rightarrow \Gamma_f ] = 
\sum_{{\cal M} | \partial {\cal M}= \Gamma_0 \cup \Gamma_f, }
{\cal A}[{\cal M}]
\ff

A theory of this form is then specified by the pair
$(G_q, {\cal A})$ of quantum groups and sets of amplitudes
on labeled four simplices.   

\section{Perturbations of histories in terms of an induced $1+1$
dimensional theory}

We now study the problem of describing a perturbation theory for
the class of theories $(G_q, {\cal A})$ based on a quantum group 
$G_q$ and amplitudes  ${\cal A}$ which we have defined.
To simplify the
presentation we will
work first 
with the simplest case in which $G_q=SU(2)_q$ and
then discuss the extension to a general $G_q$.

To define a perturbation of a history we must first define the 
perturbation
of a state $|{\cal T}, \{ j\} , \{ \mu \} \rangle 
\in {\cal H}^\Sigma$.  
There are two classes of perturbations: those that change only
the labelings, $(\{ j\} , \{ \mu \} )$ leaving the triangulation
${\cal T}$ fixed, and those that change the triangulation.
Changes in the triangulations are generated by the Pachner moves
but these are also the evolution moves.   To cleanly
separate evolution along one history from perturbations 
that take us from a history to a distinct history we consider
only the first class of perturbations.

To study these perturbations 
we  need to know how, given an initial set of consistent
labels $(\{ j\} , \{ \mu\})$ new sets
$(\{ j^\prime\} , \{ \mu^\prime \})$ may be chosen that are
both consistent and differ from the original
set by a small change.  
To solve this problem we should remember that the consistency
conditions express the gauge invariance of the theory under
the gauge group  $SU(2)_q$.  
Consistent changes in the labeling correspond to addition of Wilson
loops around closed loops on the graph.  In the dual picture in which
we are working
consistent perturbations are defined by putting a closed
loop of labels in the elementary representation
$1$ in the triangulation corresponding to the addition of a Wilson
loop to a dual graph.  
This corresponds to a loop
$\gamma \in {\cal T}$ of labels, by which we mean a sequence of 
alternating faces
$f_i$ and tetrahedra $\tau_i$,
\f
\gamma = \{ \tau_1 , f_1 , \tau_2 , f_2 , \tau_3 , ......, \tau_n , f_n \}, 
\label{eq:gamma}
\ff
where each face is between  two adjacent tetrahedra:
$f_i \in \tau_i$ and $f_i \in \tau_{i+1}$.  
We then define the new state to be
\f
|\gamma * \Psi \rangle = |{\cal T}, \{ j^\prime \} , \{ \mu^\prime \} \rangle.
\label{loop}
\ff
where the new representations and intertwiners are changed only
for the faces and tetrahedra (\ref{eq:gamma}) in $\gamma$. 

The representations are changed by 
 taking the product along each link and
intertwiner of the old label with the spin $1/2$ edge,
giving us a superpositions of new labels coming
from the decomposition of $j^\prime = 1 \otimes j$.
The change in the intertwiners is obtained
 by splitting the 4-valent
node associated to each tetrahedron along the path of
the loop $\gamma$.
To calculate the result, we use
the edge addition formula of spin networks 
(addition of angular momentum) \cite{lou-sn}. 
Let  the loop $\gamma$
cross $n$ faces and hence $n$ intertwiners, labelled by
$j_i$ and $k_i$, respectively. Call $|{\cal T},
 \{ j +\delta \} , \{ k +\rho \} \rangle$ the state where the i'th 
face that $\gamma$ crosses has been changed from $j_i$ to $j_i+\delta_i$ 
and the $i$'th intertwiner has been changed from
$k_i$ to $k_i +\rho_i$. Then the new perturbed state is 
\f
|\gamma * \Psi \rangle = \sum_{\rho_i=\pm 1}\sum_{\delta_i=\pm 1}
|{\cal T}, \{ j +\delta \} , \{ k +\rho \} \rangle\prod_{a} C(a) .
\label{pertstate}
\ff
$C(a)$ is a weight factor for each trivalent node $a$ in the splitting
of the 4-valent ones along $\gamma$.
Each $C(a)$ depends on a face spin $j$ (and the corresponding $\delta$), 
an intertwiner spin  $k$ (and the corresponding $\rho$)
 and a third spin $r$, from
an edge not crossed by $\gamma$.  Thus, 
$C(a)=C^{jkr}_{\delta \rho}$.
From recoupling theory we find that\cite{lou-sn},
\bea
C^{jkr}_{++}&=&1\\
C^{jkr}_{+-}=C^{jkr}_{-+}&=& (-1)^k {j \Theta (k+1,k,1) \over (j+1)(k+2)}
\left \{ \begin{array}{c}
	k,\ r,\ k+1\\
	j-1,\ 1,\ j
	\end{array}         
\right \}\\
C^{jkr}_{--}&=&  (-1)^{k-1}{j \Theta(k-1,k,1) \over (j+1)(k+1)}
\left \{  \begin{array}{c}
	k,\ r,\ k+1\\
	j-1, \ 1, \ j
	\end{array}          
\right \}  .
\eea
where the symbols are the quantum deformed theta and 6j 
symbols
defined in \cite{lou-sn}.
Now that we know what perturbations of states we are interested
in, let us define the corresponding perturbations of histories.

A perturbation of a history $\cal M$ should lead to a perturbation of
every state $|{\cal T}, \{ j \} , \{ k  \} \rangle$, or
spacelike slice $\Gamma$, of the history $\cal M$.  
The many-finger-time structure of the histories $\cal M$
imposes a strong constraint on the form of the perturbation of
a history, because whenever two spacelike slices overlap the
perturbations must agree on the overlap.  This means that the
perturbation of a history must be given by a two surface $S$
embedded in the history $\cal M$ in such a way that every slice
through it is a loop of the form of (\ref{loop}) that gives
a perturbation of the state corresponding to that slice.
 
We will also require our perturbations to be causal. 
Causal perturbations are those  in which the support
of the perturbation on any slice is in the causal
future of the support of the perturbation of the previous
slice. This is satisfied when (a) 
for any spacelike slice $\Gamma$,
$S \cap \Gamma$ is a closed loop $\gamma (S) \in \Gamma$, and   
(b) for any 4-simplex in $\cal M$, if $S$ includes at 
least	
one of its future tetrahedra it also includes at 
least one of its past tetrahedra (i.e. $S$ is 
timelike)\footnote{A perturbation
of a history based on a surface $S$ which is not causal cannot be assigned
a well defined effective action by the procedure described here.}.

Thus, we have two spin fields on $S$. The $\delta=\pm 1$ live
on the perturbed faces, while the $\rho= \pm 1$ live on 
the perturbed tetrahedra.  Thus, the perturbation defines a rather 
complicated
spin system on the 2-dimensional surface. The
perturbation $\cal M'$ of the history $\cal M$,
is the superposition of all the histories in 
which the $\delta $'s and $\rho$'s take all their possible
values.  

A history $\cal M$ is an amplitude from an initial spin 
network state to a final one. This amplitude is given by the product of 
the amplitudes for the 4-simplices $S_4$ that make up  the history 
\cite{fotini1},
\f
{\cal A}[{\cal M}]=\prod_{I}{\cal A}_I[j,k,c].
\label{eq:historyamp}
\ff
The amplitude $\Delta W$ of the perturbation $S$ is then given by
\f
\Delta W [{\cal M}, S] = W[{\cal M}^\prime ] - W[{\cal M}].
\ff
$\Delta W$ can be calculated as an induced amplitude from 
(\ref{eq:historyamp}) for the 4-simplices that $S$ intersects. It is 
found to be
\f
\Delta W [{\cal M}, S] = -i \ln \left \{
\sum_{\delta, \rho=\pm 1}\prod_{{S_4}\in S \cap {\cal M}}
{{\cal A}[S_4; j+\delta , k+\rho ] (\prod C_{\delta \rho} 
)   
\over  {\cal A}[S_4; j , k ]   }
\right \}
\label{p1}
\ff
with
the products taken over a set of loops in a foliation of
$\cal M$.  
 
Contributions  from 
4-simplices not in the surface cancel out.  As a result, the
action of the perturbation is given by
an effective spin system on the surface $S$ with couplings given
by the spins, intertwiners and causal structure of the
4-simplices on $S$.   That is, we may compute the
cost of the perturbation by finding the vacuum to vacuum
amplitude of the spin system whose classical action is given
by 
\f
S^{eff}[\delta, \rho ] = \sum_{S_4\in S \cap {\cal M}} \ln \left \{
{{\cal A}[S_4; j+\delta , k+\rho ] (\prod C_{\delta \rho} 
)   
\over  {\cal A}[S_4; j , k ]   }
\right \} 
\label{p2}
\ff
Thus, there is a 2-dimensional field theory
associated with the perturbations of a history.   This is the
main result of this paper.  In the following sections we discuss
its implications.

\section{The classical limit and string theory}

So far we have not invoked any assumption about the classical
limit.  The basic idea we will explore now is that 
if the full quantum theory of gravity has a classical limit,
its perturbations should also have a classical limit, which can
be studied by analyzing the 2-dimensional
system just derived.
It is a nontrivial task to analyze the critical behavior of this 
2-dimensional system, but at the same time it is likely to be a far
easier task than the analysis of the full 4-dimensional theory.
As a first step, one can
argue to a useful conclusion as follows.

Let us assume that a particular history $\cal M$ 
and perturbation $S$ is chosen such that

\begin{enumerate}

\item{} There is a smooth 4-dimensional spacetime and metric,
$(\widetilde{\cal M}, \tilde{g}_{ab})$ and an embedding map
$e$ that embeds  $\cal M$  in $\widetilde{\cal M}$ by assigning to
the tetrahedra $\tau$ of $\cal M$ events, $e(\tau)$ in $\widetilde{\cal M}$,
such that (a) the causal structure of the events $e(\tau)$ agrees
with the causal set structure of $\cal M$, and 
(b)  for each spacelike slice $\Sigma$ in $\cal M$ there is a 
spacelike slice $\widetilde\Sigma$ in $\widetilde M$ such 
 that the
areas of large surfaces and volumes of large regions 
computed from either the smooth euclidean metric $h_{ab}$ on
$\widetilde\Sigma$ or
the labelings of faces and
tetrahedra in $\Sigma$ agree up to small errors
(where small and large are defined in Planck units).
We then  say that the spacetime 
$(\widetilde{\cal M}, \tilde{g}_{ab})$ is the ``classical limit"
of the discrete causal history $\cal M$.  

\item{} $g_{ab} = \eta_{ab}$, the Minkowski metric. 

\item{}The causal structures and labelings of the 4-simplices of
$\cal M$ are distributed randomly with respect to $g_{ab}$, so that
the distribution is invariant under a Poincar\'e transformation.

\item{}Corresponding to the surface $S \in {\cal M}$ there is
a timelike cylindrical surface $\widetilde{S} \in \tilde{\cal M}$.  
$\widetilde{S}$ has
large area and small extrinsic curvature.

\end{enumerate}

Under these assumptions the action of the 
perturbation becomes
\f
\Delta W[{\cal M}, S] = w N_4 [S \cap {\cal M}]
\ff
where $N_4 [S \cap {\cal M}]$ is the number of 4-simplices 
 $S$ crosses and  $\cal M$ and $w$ is the average value 
contributed to $\Delta W$  by one 4-simplex, when averaged over the 
different
types and labelings that may appear.

However, by Poincar\'e invariance, the number $N_4 [S \cap {\cal M}]$
can only be proportional to the area $A[\widetilde{S}, g_{ab} ]$
of $\widetilde S$  computed
from the spacetime metric $g_{ab}$, as that is the unique additive
Poincar\'e invariant measure of the surface.  Thus,
\f
N_4 [S \cap {\cal M}]= {c \over l_{Pl}^2} A[\tilde{S}, g_{ab} ],
\ff
which gives $\Delta W$ as 
\f
\Delta W[{\cal M}, S] = {c w \over l_{Pl}^2} A[\tilde{S}, g_{ab} ].
\label{p3}
\ff
Thus we find that the action of the perturbation
is proportional to the area of a 2-dimensional timelike
surface. It is intriguing that this agrees
 to leading order with the Nambu action for the bosonic string.  We may note
also that the string scale $\alpha^\prime = l^2_{Pl} /cw$ is 
computable in terms of the fundamental theory.

Of course, this is not enough to allow us to conclude that
the perturbation theory given by (\ref{p2}) in fact becomes
a consistent string theory under the assumptions we have
indicated.  It may just be the case that the effective
action (\ref{p3}) describes an effective, non-critical string
theory as in $QCD$.  In fact, this is likely to be true in the
case we have just considered.  There is no consistent bosonic
string theory based on the Nambu action, without additional
degrees of freedom, in $3+1$ dimensions.  We may note that this
is consistent with the evidence that quantum general relativity
in $3+1$ dimensions has no massless 
particles\cite{trouble1,trouble2}.

\section{Critical behavior and string theory}

The question is then whether there is some extension of
the formalism we considered in the last section in which
$SU(2)_q$ is replaced by a general group or supergroup
$G_q$ which has good critical behavior for some choice
of amplitudes ${\cal A}[j,k,c]$.  This means that the theory will
have a semi-classical  limit in which histories $\cal M$ may
be described by a classical spacetime $(\tilde{\cal M},\tilde{g})$
with various quantum fields $\hat{\phi}$ living on it.  These
must include massless particles corresponding to the graviton, as
well as possibly chiral fermions and gauge fields.  In the case that the
classical limit $(\tilde{\cal M},\tilde{g})$ of the history
$\cal M$ is Minkowski spacetime, the perturbation theory
must also be stable, otherwise we do not have a good theory of
quantum gravity.

The same technique used in section 4 can be used for a 
general $G_q$  to obtain a $1+1$ dimensional discrete
field theory.  The spin variables $\delta_i$ and $\rho_i$
are replaced by more complicated variables which describe
the multiplication of the representations and intertwiners of
$G_q$ by one of the elementary representations.  The result
will be a generalization of (\ref{p2}).   

There is a simple argument that this extension of (\ref{p2})
must include the degrees of freedom of a consistent string theory.
This depends only on the assumptions that the semi-classical
limit of the theory $(G_q, {\cal A})$ has massless gravitons
and is stable around Minkowski spacetime.

The degrees of freedom in the theory (\ref{p2}) describe the
possible gauge invariant perturbations of a history $\cal M$
that has a good classical limit.  Included in
the perturbation theory must be the massless graviton and other massless
degrees of freedom.  This means that the $1+1$ theory
defined by the extension of (\ref{p2}) must itself have
massless modes which correspond to 
the graviton and other massless modes. This is because the
$1+1$ theory has been defined so that it includes
all the weakly coupled long ranged modes described by the full theory. 
These massless modes
propagate along the lightcone of the
two surface, which is induced from the 
causal structure of the history $\cal M$, which by construction
must agree with the lightcone of
the spacetime $(\tilde{\cal M},\tilde{g})$. It follows that there
must be a sector of the theory given by the extension of
(\ref{p2}) which is conformally invariant, which means it
describes a consistent perturbative string theory.

Moreover, this string theory must be stable when the
background $(\tilde{\cal M},\tilde{g})$ is Minkowksi 
spacetime as this follows from the assumption that the
perturbation theory must be stable.  

On the other hand, suppose that, as is likely the case with
pure general relativity, (\ref{p2}) or its extensions do not
describe a consistent  perturbative string theory
in $3+1$ dimensions.  Then the theory $(G_q, {\cal A})$
does not have a perturbation theory which contains
massless particles.  The same is true for the issue of
stability; if the theory defined by the extension of
(\ref{p2}) is not stable around Minkowksi spacetime
then there are small perturbations of the full 
non-perturbative theory that destabilize Minkowski
spacetime.  This means that it fails as a possible
quantum theory of gravity.  So we see that a necessary
condition for a non-perturbative quantum theory of gravity
$(G_q, {\cal A}) $ to have a good continuum limit that
reproduces classical general relativity plus quantum
field theory is that the $1+1$ theory described by
(\ref{p2}) has a sector that corresponds to a consistent
stable perturbative string theory.

We may note that there are many stable, consistent perturbative
string theories in $3+1$ dimensions.  These can be constructed
by compactification or by adding degrees of freedom to the 
worldsheet.  It is also known that discrete $1+1$ models may
give rise to consistent perturbative string 
theories\cite{ks,matrix,dvv}.  It will then be interesting to 
investigate whether there exists a choice of $(G_q,{\cal A})$
such that the extension of (\ref{p2}) leads to one of these
consistent $3+1$ dimensional string theories.  If so, this will
establish a relationship between non-perturbative quantum
gravity  and perturbative string theory.

\section*{ACKNOWLEDGEMENTS}

The results reported here grew out of joint work with Fotini
Markopoulou and I must thank her for many suggestions
and criticisms which improved my understanding of them.
I would also like to thank  Louis Crane, Sameer Gupta, 
Stuart Kauffman and  Carlo Rovelli
for many discussions about this direction of work.
This work was supported by
NSF grant PHY-9514240 to The Pennsylvania State
University and a NASA grant to the Santa Fe Institute.

\end{document}